\documentstyle[prl,preprint,aps]{revtex}
\begin{document}
\preprint{mpifkf-01.94}
\draft
\title{Localization and dynamic persistent currents in long cylinders}
\author{S. Kettemann$^{\left( 1 \right)}$\cite{weiz}, K. B. Efetov$^{(1,2)}$}
\address{$\left( 1\right) $ {\it Max-Planck Institut f\"urFestk\"orperforschung
}%
Heisenbergstrasse1, D-70569 Stuttgart, Germany\\$\left( 2\right) $
Landau Institute for Theoretical Physics, Moscow, Russia}
\date{\today }
\maketitle

\begin{abstract}
A dynamic response to a magnetic field in a long disordered cylinder is
considered. We show that, although at high frequencies conduction is
classical in all directions, the low frequency behavior corresponds to
localization in the longitudinal direction and to a diamagnetic dynamic
persistent current in the transversal one. The current density does not
vanish even in the limit of the infinitely long cylinder. Being of a purely
dynamic origin the current can be destroyed by inelastic scattering but at
low temperatures the decay time can be very large.
\end{abstract}
\pacs{72.10.Bg, 05.30.Fk, 71.25.Mg, 75.20.En}
\newpage
The recent measurement of the magnetization of an array of $10^7$ isolated
Cu-rings \cite{p0} attracted a lot of attention to the problem of persistent
currents in disordered systems. The circumference of these rings was much
larger than the elastic mean free path and therefore at short distances
electron motion was diffusive. Considerable effort has been taken in order
to build a proper theory of the observed effect in the diffusive regime.
Calculation of the persistent current by averaging over disorder at fixed
Fermi energy gives in this regime an exponentially small amplitude. In order
to obtain a non vanishing effect one should either take into consideration
electron-electron interactions \cite{k1} or calculate with a fixed number of
particles \cite{r9,r10,r11}. In both   cases it is essential that all the
sizes of the sample are small. It is clear that for a macroscopic object the
difference between results for the canonical and grand canonical ensembles
cannot be important and therefore one should not hope to have a
thermodynamically stable persistent current in such a case.

However, does it mean that there is no possibility to have a long living
macroscopic current in a macroscopic piece of a disordered metal? Of course,
if such a current did not correspond to the minimum of the free energy
different types of inelastic scattering would kill the effect. Nevertheless,
the corresponding decay times would be very large and strongly dependent on
temperature which makes a discussion of such a possibility quite interesting.

In order to understand better what type of effect we want to consider below
let us imagine first a very long clean metallic cylinder with the length $%
L_z,$the circumference $L$ and the thickness of the walls $d$. Applying a
magnetic field $H$ along the cylinder one induces currents. In the ballistic
regime the cylinder is an ideal diamagnet and the induced total current $I_d$
is equal to
\begin{equation}
\label{e1}I_d=dL_zLK_dH/4\pi c,\qquad K_d=-ne^2/m.
\end{equation}
In Eq.(\ref{e1}) $m$ is the electron mass, $e$ is the electron charge, $n$
is the electron density, $c$ is the light velocity. The current $I_d$ can
exist at very low temperatures for a very long time but it does not
correspond to the free energy minimum and must finally decay.
Experimentally, the current can be obtained first cooling down the cylinder
and then applying the magnetic field. The inverse procedure would lead to a
thermodynamic current which can be much smaller. The averaged thermodynamic
persistent current $I_{th}$ for clean cylinders was calculated in ref.\cite
{k2} and can be estimated as
\begin{equation}
\label{e3}I_{th}\sim \left| I_d\right| /(p_0L)^{1/2},
\end{equation}
where $p_0$ is the Fermi momentum. Eq.(\ref{e3}) shows that the
thermodynamic persistent current can be much smaller than the dynamic one
which, of course, is not unexpected. The difference between the dynamic and
thermodynamic currents is due to existence of degenerate states. If one
represents energy levels by functions of the magnetic field the degeneracies
correspond to level crossing. In the diffusive regime $p_{0_{}^{}}^{-1}\ll
l\ll L,$where $l$ is the mean free path, the thermodynamic current is
proportional to $\exp (-L/l)$ \cite{k2} and is negligibly small. In this
Letter we calculate the corresponding dynamic current and  show that it
can be quite noticeable. For simplicity we consider a standard model of non
interacting electrons moving in a random potential. Calculating the dynamic
current we use standard formulae for linear response. We assume that the
external magnetic field $\tilde H$ has the form
\begin{equation}
\label{e5}^{}\tilde H=H+H_\omega \cos (\omega t),
\end{equation}
where $H$ is the static component which can be arbitrary. The second term in
Eq.(\ref{e5}) describes the oscillating part of the magnetic field which
induces an electric field directed along the circumference and,
correspondingly, an oscillating current. We assume that the amplitude $%
H_\omega $ of the oscillating part of the magnetic field is small and
calculate the induced current using the conventional linear response theory.
The density of the oscillating current $j_{\omega \text{ }}$can be written
as
\begin{equation}
\label{e6}j_\omega =c^{-1}K(\omega )A_\omega,
\end{equation}
where $K\left( \omega \right) $ is the current-current correlation function
and $A_{\omega \text{ }}$is the vector potential corresponding to the
magnetic field $H_\omega $ in Eq.(\ref{e5}). We use the gauge with the
vector potential directed along the circumference. Using the retarded $%
G_\epsilon ^R\left( r,r^{^{\prime }}\right) $ and advanced $G_\epsilon
^A\left( r,r^{^{\prime }}\right) $ Green functions we can write the function
$K\left( \omega \right) $ as follows
\begin{equation}
\label{e7}K\left( \omega \right) =K_1\left( \omega \right) +K_2\left( \omega
\right),
\end{equation}
\begin{equation}
\label{e8}K_1\left( \omega \right) =(e^2/i\pi )\int \left[ n\left( \epsilon
-\mu \right) -n\left( \epsilon -\omega -\mu \right) \right] \left\langle
\hat v_rG_\epsilon ^R\left( r,r^{^{\prime }}\right) \hat v_{r^{^{\prime
}}}G_{\epsilon -\omega }^A\left( r^{^{\prime }},r\right) \right\rangle
dr^{^{\prime }}d\epsilon,
\end{equation}
\begin{equation}
\label{e9}
\begin{array}{c}
K_2\left( \omega \right) =(e^2/i\pi )\int n\left( \epsilon -\mu \right)
\left\langle \left[
\hat v_rG_{\epsilon +\omega }^R\left( r,r^{^{\prime }}\right) \hat v%
_{r^{^{\prime }}}G_\epsilon ^R\left( r^{^{\prime }},r\right) -\hat v%
_rG_\epsilon ^A\left( r,r^{^{\prime }}\right) \hat v_{r^{^{\prime
}}}G_{\epsilon -\omega }^A\left( r^{^{\prime }},r\right) \right]
\right\rangle dr^{^{\prime }}d\epsilon + \\  \\
(e^2/i\pi m)\int n\left( \epsilon -\mu \right) \left\langle \left[
G_\epsilon ^R\left( r,r\right) -G_\epsilon ^A\left( r,r\right) \right]
\right\rangle d\epsilon,
\end{array}
\end{equation}
where $\hat v_r=\left( 1/m\right) \left[ -i\nabla -\left( e/c\right)
A\right] $ is the component of the velocity operator perpendicular to the
axis of the cylinder, $A$ is the vector potential corresponding to the
static component $H$ of the magnetic field, $n\left( \epsilon \right) $ is
the Fermi distribution, and the angular brackets stand for averaging over
impurities. The terms $K_1\left( \omega \right) $ and $K_2\left( \omega
\right) $ have completely different origins. In the limit $\omega
\rightarrow 0$ the term $K_2\left( \omega \right) $ can be obtained from the
free energy by differentiating at a fixed chemical potential over the vector
potential and, as we have mentioned, can be much smaller than the first one.
In the diffusive regime the average over impurities of a function
containing a product of
only retarded or only advanced Green functions can be substituted by the
product of the corresponding averaged Green functions $\left\langle
G^{R,A}\right\rangle .$ In the momentum representation this is given by
\begin{equation}
\label{e10}\left\langle G_\epsilon ^{R,A}\right\rangle =\left( \epsilon
-\epsilon \left( \hat p\right) \pm i/2\tau \right) ^{-1},
\end{equation}
where $\tau $ is the mean free time. Substituting Eq.(\ref{e10}) into Eq.(%
\ref{e9}) and integrating the first term in Eq.(\ref{e9}) over momentum by
parts one can see that in the limit $\omega \rightarrow 0$ it cancels the
second one.
Here, we consider the  situation when the magnetic field is concentrated
 in the cavity of the cylinder.  Otherwise, we would have to  sum over
the Landau levels  and would obtain  that the quantity $K_{2\text{ }}$%
gives additionally the Landau diamagnetism.

In the following we concentrate on calculation of the quantity $K_1\left(
\omega \right) $ . Before carrying out explicit calculations in the
diffusive regime let us demonstrate how one can calculate this quantity in
the ballistic one. Using the representation $^{}$
\begin{equation}
\label{e11}G_\epsilon ^{R,A}\left( r,r^{^{\prime }}\right) =\sum_\alpha
\frac{\varphi _\alpha \left( r\right) \varphi _\alpha ^{*}\left( r^{^{\prime
}}\right) }{\epsilon -\epsilon _\alpha \pm i\delta },
\end{equation}
and taking the limit $\omega \rightarrow 0$ we obtain from Eq.(\ref{e8})
\begin{equation}
\label{e12}\lim _{\omega \rightarrow 0}K_1\left( \omega \right)
=-(2e^2/V)\sum_{\epsilon _\alpha =\epsilon _\beta }\left| \hat v_{\alpha
\beta }\right| ^2\delta \left( \epsilon _\alpha -\mu \right),
\end{equation}
where $\mu $ is the chemical potential, $V$ is the volume.

In the clean macroscopic cylinder under consideration the eigenstates are
plane waves and the matrix elements in Eq.(\ref{e12}) can easily be
calculated. All sums are to be substituted by an integral over momenta and
we can see that $\lim _{\omega \rightarrow 0}K_1\left( \omega \right) =K_d$ .

The classical limit in the diffusive regime corresponds to high frequencies $%
\omega $. In this limit one can substitute the Green functions in Eq.(\ref
{e8}) by their averages Eq.(\ref{e10}) and then  calculate the integral
over momenta. As the result one has
\begin{equation}
\label{e13}K_1(\omega )=-i\omega \sigma _0,
\end{equation}
where $\sigma _0=e^2\tau n/m$ is the Drude conductivity.

The limit of low frequencies requires more sophisticated methods of
calculation. In the diffusive regime the supersymmetry method developed by
one of the authors\cite{susy} is by now the only possibility to obtain
results analytically. This method has been used to study localization in
long disordered wires \cite{susy,k4}. As we will see below the problem of
localization is closely related to \cite{ref13,dyn1} the problem of the
dynamic persistent currents. The responses both in the longitudinal and
transversal directions can be reduced to calculation of functional integrals
over supermatrices $Q$. Calculating a correlation function for a model of a
wire corresponding to a longitudinal response for the cylinders it was shown
in Refs.\cite{susy,k4} that the system has dielectric properties in the
longitudinal direction. Here, we present  a calculation of the transversal
response
describing the persistent currents. In Refs.\cite{ref13,dyn1} a general
expression for the function $K_1\left( \omega \right) $ has been presented
in terms of integrals over supermatrices $Q$ with a free energy functional
of the non linear supersymmetric $\sigma $ -model. This expression has the
following structure
\begin{equation}
\label{a1}K_1\left( \omega \right) =-i\omega \left\langle B\left( Q\right)
\right\rangle _Q,
\end{equation}
where $B\left( Q\right) $ is a sum of products of elements of the supermatrix
$Q$.

In Eq. (\ref{a1}) the notation $\left\langle \ldots \right\rangle _Q$ stands
for the functional average
\begin{equation}
\label{e19}\left\langle \ldots \right\rangle_Q =\int \left( \ldots \right)
\exp \left( -F\left[ Q\right] \right) DQ,
\end{equation}
where $F\left[ Q\right] $ is the free energy functional of the
non-linear supersymmetric $\sigma $-model.

The form of the function $R\left( \omega \right) $
which denoted in
 Refs.\cite{ref13,dyn1} the average $\left\langle B\left( Q\right)
\right\rangle _Q$ remains unchanged for the sytem studied, here.
While the  dynamic response for the
problem of persistent currents in mesoscopic rings is calculated using the
zero dimensional version of the $\sigma $-model\cite{ref13,dyn1},
for the problem of localization in wires one has to take the one dimensional
version of the $\sigma $-model \cite{susy,k4}. Alhtough
for the cylinder in the diffusive regime
 the one dimensional $\sigma $-model is valid, let us consider a more
general model of a stack of small metallic rings. Changing the probability
of tunneling from ring to ring we can describe the crossover from the case
of the homogeneously weakly disordered cylinder to the case of isolated
rings. For such a model the derivatives in the longitudinal
coordinate are substituted by finite differences. Then the free energy $%
F\left[ Q\right] $ takes the form
\begin{equation}
\label{bethe}F[Q]=-\sum_{ij}J_{ij}STr\left[ Q_iQ_j\right] +\sum_iF_0\left[
Q_i\right],
\end{equation}
\begin{equation}
\label{bethe1}F_0\left[ Q\right] =\frac{\pi \nu Sh}8\sum_iSTr\left[ -{\cal D}%
_0\left( \frac ec{\ }{\bf A}[Q_i,\tau _3]\right) ^2+2i\omega \Lambda
Q_i\right],
\end{equation}
where $S=Ld$ is the cross section of the cylinder, $h$ is the height of each
ring, $i$ and $j$ enumerate the rings in the stack, $D_0$ is the classical
diffusion coefficient. ( We remind that $Q^2=1$). The function $F_0\left[
Q\right] $ in Eqs.(\ref{bethe},\ref{bethe1}) describes electron motion
inside the rings, the first term in Eq.(\ref{bethe}) stands for coupling
between the rings. We assume that only nearest neighbors interact,so that $%
J_{ij}=J$ for the neighbors,and $J_{ij}=0$ otherwise. The limit $J=0$
corresponds to the stack of the isolated rings. In the limit $J\gg 1$ the
model on the lattice Eq.(\ref{bethe}) can be substituted by the continuous
one and we return to the one dimensional $\sigma $-model. The notation $STr$
stands for the Supertrace, the definition of the matrices $\tau _3$ and $%
\Lambda $ can be found in previous works \cite{susy,ref13,dyn1}.

To recover the case of the homogeneously weakly disordered cylinder one
should relate $J$ to $D_0$ as
\begin{equation}
\label{e20}J=\pi \nu SD_0/8h.
\end{equation}
The main contribution for the transversal current when calculating the
integral Eqs.(\ref{a1}, \ref{e19}) comes from terms containing products of
two matrix elements of $Q$ taken at the same point ( terms $R_1\left( \omega
\right) $ and $R_3\left( \omega \right) $ in Refs.\cite{ref13,dyn1}) and the
substitution of the weakly homogeneous cylinder by the stack of the rings
does not change the form of these terms.

Due to the one dimensionality of the model Eq.(\ref{bethe}) one can use the
transfer matrix technique. Corresponding partial differential equations have
been written in Refs.\cite{susy,k4} for the continuous $\sigma $-model. For
the model on the lattice, Eq.(\ref{bethe}), the corresponding recurrence
equation is an integral. Analogous integral equations were written in
Refs.\cite
{m1,m3} for the model on the Bethe the lattice and also for one dimensional
chains. Repeating the main steps of these works we reduce calculation of
integrals over $Q_i$ for all cites $i$ to one integral over $Q$. The
corresponding changes can be done substituting Eq.(\ref{e19}) by
\begin{equation}
\label{e21}\left\langle \ldots \right\rangle _Q=\int \left( \ldots \right)
\Psi ^2\left( Q\right) \exp \left( -F_0\left( Q\right) \right) dQ.
\end{equation}
The function $\Psi \left( Q\right) $ in Eq.(\ref{e21}) satisfies the equation
\begin{equation}
\label{e22}\Psi \left( Q\right) =\int \exp (2JSTrQQ^{^{\prime }})\exp \left(
-F_0\left( Q^{^{\prime }}\right) \right) \Psi \left( Q^{^{\prime }}\right)
dQ^{^{\prime }}.
\end{equation}
The solution of the Eq.(\ref{e22}) depends on the vector potential $A$
entering $F_0$ Eq.(\ref{bethe1}). In principle, it can be solved, at least
numerically, for arbitrary magnetic fields using the parameterization
proposed recently \cite{iida} but we will present results only in the limits
of zero and high magnetic fields. These limits correspond to the orthogonal
and unitary ensembles. In both  cases one can omit the first term in Eq.(%
\ref{bethe1}) because in the unitary case the supermatrix $Q$ commutes with
the matrix $\tau _3$ \cite{susy}. Due to the symmetry of the free energy the
function $\Psi \left( Q\right) $ depends only on the variables $\lambda $ and
$\lambda _{1,2}$ parameterizing the supermatrix $Q$ \cite{susy}. Therefore,
in the integral over $Q$ in Eq.(\ref{a1}) one can integrate first over all
other variables reducing thus the integral to an integral over $\lambda $
and $\lambda _{1,2}$. At high frequencies deviations of the supermatrix $Q$
from $\Lambda $ are small and one comes to Eq.(\ref{e13}). In this limit the
conduction is classical in all directions.

Calculations for arbitrary frequency are most simple for the unitary
ensemble.
Corresponding calculations are not very different from those performed in
Refs.\cite{ref13,dyn1} and we obtain
\begin{equation}
\label{e23}K_1^{unit}\left( \omega \right) =-i\omega \sigma _0\left\{ 1+%
\frac 12\int_{-1}^1\int_1^\infty \exp \left[ \left( i\pi \omega /\Delta
-\delta \right) \left( \lambda _1-\lambda \right) \right] \Psi ^2\left(
\lambda ,\lambda _1\right) d\lambda d\lambda _1\right\},
\end{equation}
where $\Delta $ $=\left( \nu Sh\right) ^{-1}$is the mean level spacing in
one ring in the stack and $\delta \rightarrow 0$.

Non zero persistent currents are possible if at low frequencies $\omega $
the second term in the brackets in Eq.(\ref{e23}) is proportional to $1/\omega
$. In the
most interesting limit $\omega \rightarrow 0$ all calculations when solving
Eq.(\ref{e22}) become more simple because the main contribution in all
integrals over $\lambda _1$comes from $\lambda _1\sim 1/\omega $. Then the
function $\Psi \left( Q\right) $depends only on one variable $z=2\omega
\lambda _1$and we obtain for the response $K_1\left( \omega \right) $
\begin{equation}
\label{e24}K_1^{unit}\left( 0\right) =-\sigma _0\Delta _{eff}\left( J\right)
/\pi,
\end{equation}
where
\begin{equation}
\label{e25}\Delta _{eff}\left( J\right) =\Delta \int_0^\infty \exp \left(
-z\right) \Psi _J^2\left( z\right) dz,
\end{equation}
is a non trivial function of the coupling $J$ between the rings. This
function is known numerically for arbitrary $J$.\cite{m3} In the limit $%
J=0$ the function $\Psi =1$ and
\begin{equation}
\label{e26}\Delta _{eff}\left( 0\right) =\Delta.
\end{equation}
In the opposite limit $J\gg 1$ the function $\Psi \left( z\right) $ is the
solution of the differential equation
\begin{equation}
\label{e27}zd^2\Psi /dz^2-16J\Psi =0.
\end{equation}
Solving Eq.(\ref{e27}) and calculating the integral Eq. (\ref{e25}) we find
\begin{equation}
\label{e28}\Delta _{eff}\left( J\right) =\Delta /\left( 96J\right) ^{}.
\end{equation}
As we have mentioned above one has localization in the longitudinal
direction. In the unitary ensemble the localization length $L_c$ was
calculated in Ref.\cite{susy} and can be related with the help of
Eq.(\ref{e20}) to the coupling $J$ as $L_c=16Jh$. In this limit the function
$\Delta _{eff}\left( J\right) $ can be rewritten as
\begin{equation}
\label{e29}\Delta _{eff}\left( J\gg 1\right) =\left( 6\nu SL_c\right) ^{-1}.
\end{equation}
We see from Eq. (\ref{e29}) that in the limit of large $J$ the response $%
K_1^{unit}\left( 0\right) $ Eq. (\ref{e24}) looks as if the cylinder
consisted of rings with the height of the order of $L_c$. Compairing Eqs. (%
\ref{e13} and \ref{e24}) we see that the characteristic frequency of the
crossover from the quantum to the classical regime is of the order $\Delta
_{eff}$. In the model of the weakly homogeneously disordered cylinder we can
see with the help of Eq. (\ref{e20}) that the response $K_1^{unit}\left(
0\right) $ does not depend on the disorder. In this limit it is very small.
Decreasing the coupling $J$ makes the localization length shorter which
leads to a larger value of the response.

Analogous calculations can be carried out for the orthogonal ensemble
corresponding to the zero static component $H$ of the magnetic field Eq. (%
\ref{e5}). In this case we obtain as in Refs.\cite{ref13,dyn1}
\begin{equation}
\label{e30}K_1^{orth}\left( 0\right) =0.
\end{equation}
Changing the static component $H$ of the magnetic field we can have a
crossover between Eqs.(\ref{e30}) and (\ref{e24}). The characteristic flux $%
\phi _c$ of this crossover is of the order of $\phi _0(\Delta
_{eff}/E_c)^{1/2}$, where $E_c=\pi ^2D_0/L^2$ is the Thouless energy, $\phi _0$
is the flux quantum. The whole dependence of the response on the flux $\phi $
is periodic with the period $\phi _0/2$.

Adding magnetic or spin-orbit impurities changes Eqs. (\ref{e24}, \ref{e30}%
). The system with the magnetic impurities corresponds to the unitary
ensemble. One can use as before Eq. (\ref{e24}) provided $\Delta
_{eff}\left( J\right) $ is substituted by $\Delta _{eff}\left( 2J\right) /2.$
If the magnetic impurities are absent spin-orbit ones lead  to a
different function $\tilde \Delta _{eff}\left( J\right) $. However, this
difference is only numerical and does not change the sign of the response
which is in all cases diamagnetic.

In conclusion, we showed that a magnetic field applied to a disordered
cylinder parallel to the axis induces a macroscopic diamagnetic current. In
the absence of inelastic scattering which can correspond to low temperatures
this current can live for a very long time. The current density remains
finite even in the limit of an infinitely long cylinder. At the same time
the longitudinal response shows a dielectric behavior usual for disordered
wires. In fact, the shorter the localization length in the longitudinal
direction is the larger is the current in the transversal one. In the limit of
high frequencies or short inelastic times the transport in the model under
consideration is classical in all directions and is described by the
conventional Ohm law.

We are grateful to V. Falko for critical reading of the manuscript.

\end{document}